\documentclass[twocolumn]{aastex631}
\usepackage{amsmath,amstext}
\usepackage[T1]{fontenc}
\usepackage{apjfonts} 
\usepackage{multirow}
\usepackage[figure,figure*]{hypcap}
\usepackage{booktabs,caption}
\usepackage{hyperref}
\usepackage{subcaption}
\usepackage{mwe}

\begin{document}

\defcitealias{Inoue14}{I14}
\defcitealias{Bassett21}{B21}
\defcitealias{Kannan22}{K22}

\title{The leakage of Lyman-continuum photons from a major merger at $z\sim1$}

\author[0009-0003-8568-4850]{Soumil Maulick}
\affil{Inter-University Centre for Astronomy and Astrophysics, Ganeshkhind, Post Bag 4, Pune 411007, India}

\author[0000-0002-8768-9298]{Kanak Saha}
\affil{Inter-University Centre for Astronomy and Astrophysics, Ganeshkhind, Post Bag 4, Pune 411007, India}

\author[0009-0008-2970-9845]{Manish Kataria}
\affil{Inter-University Centre for Astronomy and Astrophysics, Ganeshkhind, Post Bag 4, Pune 411007, India}

\author[0000-0002-8505-4678]{Edmund Christian Herenz}
\affiliation{Inter-University Centre for Astronomy and Astrophysics, Ganeshkhind, Post Bag 4, Pune 411007, India}



\begin{abstract}
We report the detection of Lyman-continuum (LyC) photons from a massive interacting system at $z=1.097$ in the Hubble Ultra Deep Field. The LyC detection is made in the far-ultraviolet F154W band of the UVIT telescope onboard AstroSat. Both JWST and HST imaging of the system reveal signs that it is a likely merger. In particular, high-resolution imaging in the JWST bands reveals an infrared luminous object within the system that is faint in the bluer HST bands. The ionized-gas kinematics from the MUSE-UDF data supports the merger hypothesis.
We estimate that the entire system is leaking more than $8 \%$ of its ionizing photons to the intergalactic medium. The SED-derived stellar masses of the two components indicate that this is a major merger with a mass ratio of ${1.13 \pm 0.37}$. This detection hints at the potential contribution of massive interacting systems at higher redshifts, when major mergers were more frequent, to the ionizing budget of the universe.

\end{abstract}

\keywords{Reionization, Ultraviolet astronomy, galaxies: interactions}


\section{Introduction}
The ionization of neutral hydrogen in the Universe is thought to have been driven by star-forming galaxies during the Epoch of Reionization (EOR; \cite{Stark16,Robertson22}). Directly observing the ionizing radiation, (Lyman-continuum photons, hereafter LyC, $\lambda < 912 \text{\AA}$) from these galaxies during this epoch is extremely difficult because of the significant optical depth of the inter-galactic medium (IGM) at these redshifts \citep{Madau95,Inoue14}. 
\par At lower redshifts, where the IGM opacity does not pose that big of a challenge, galaxies emitting LyC radiation, called 'LyC leakers', have been detected in a range of redshifts, at $z<0.4$ (eg: \cite{Izotov18,Izotov21,Flury22a}), at $1<z<2$ (e.g., \cite{Saha20,Dhiwar24}) and at $z>2$ (e.g., \cite{Vanzella16,Steidel18,RiveraThorsen19,MarquesChaves22}). For a more comprehensive list of individual LyC detections we refer the reader to \cite{Kerutt23}. Studying such galaxies can offer insights into the processes that facilitate LyC leakage, and the escape of LyC photons is quantified by the escape fraction ($f_{\rm{esc,LyC}}$). 
However, the physical processes that govern this quantity, and their relation to the nature of the LyC emitting galaxy are not clear. 
\par Mechanisms of LyC escape based on photo-ionization modeling have been proposed such as the ionization-bounded and density-bounded models of \cite{Zackrisson13,Paswanetal2022}. These can guide observations, where by using emission and absorption lines one may probe outflows that relate to feedback-driven density bounded regions that may facilitate the escape of ionizing photons \citep{Perrotta21}.
Alternatively, the ionization state of the ISM can also be used to pick out LyC leaker candidates \citep{Wang19,Wang21,Nakajima20,Flury22}. Confirmed LyC leakers in the redshift range, $z\sim 0.3-0.4$, are mostly low-mass ($\text{log}_{10}(\text{M}_{*}/\text{M}_{\odot})<10$), compact and highly star-forming systems with large H$\beta$ equivalent widths and high $[\text{O III}] \lambda 5007/ [\text{O II}] \lambda 3727$  (O32) ratio \citep{Izotov2016,Izotov18,Izotov21}. In the local universe, high-resolution imaging can help ascertain the possible mechanisms of LyC escape. One famous example is that of the LyC leaker, Haro 11 \citep{Bergvall06}, where high-resolution imaging of the system in multiple bands \citep{Adamo10,Ostlin15,Ostlin21,Prestwich15,Komarova24} reveals that the system is disrupted because of merger-driven interactions and is comprised of multiple star-forming knots. Recent 21cm HI imaging \citep{LeReste23} of Haro 11 has provided support for the displacement of neutral hydrogen gas from the center of LyC production, which should act favorably for facilitating LyC escape. These observations provide a potential link between LyC escape and galaxy mergers. HI imaging of Ly$\alpha$ emitters in the local universe has also revealed that some of these systems have either an extended or disturbed HI morphology due to possible interactions with companions, which in turn could facilitate the escape of LyC photons \citep{Cannon04,Purkayastha22}. Additionally, mergers can trigger starbursts \citep{Barnes91,Teyssier10} which in turn enhance the intrinsic LyC photon production rate within these systems.\\
In this work, we report the detection of LyC photons from a major merger at $z=1.097$. We present high-resolution HST and recent JWST imaging of this massive ($\text{log}_{10}(\text{M}_{*}/\text{M}_{\odot})\gtrsim 10$) interacting system. We describe its morphology, analyze the kinematics of its ionized gas and provide constraints on the LyC escape fraction, before briefly discuss the implications of our finding.
All magnitudes quoted are in the AB system \citep{Oke83}. Throughout this work, we adopt the concordance $\Lambda$CDM cosmological model, with $H_0=70 \text{km}\text{s}^{-1}\text{Mpc}^{-1}$, $\Omega_m =0.3$, $\Omega_{\Lambda}=0.7$.

\begin{table}
\centering
\nolinenumbers
\caption{Properties of AF13753 and its individual components G1 and G2. The CANDELS and MUSE IDs of the object are from the \cite{Whitaker19} and \cite{Bacon23} catalogs respectively. All reported line fluxes are in units of $10^{-17}\text{erg}\ \text{s}^{-1}\ \text{cm}^{-2}$ and have been corrected for Galactic and internal extinction. $\text{SFR}_{\text{SED,10Myr}}$ corresponds to the star-formation rate computed in the past 10 Myrs for a Salpeter IMF, for the best fit CIGALE SED. The $\text{SFR}_{\text{H}\alpha}$ is calculated using the \cite{Kennicutt98} relation calibrated for a Salpeter IMF. $f^{\rm{H}\alpha}_{\rm{esc,LyC}}$ corresponds to the absolute escape fraction estimated using the dust-corrected H$\alpha$ emission line and is also corrected for IGM absorption. $\tilde{f}^{\rm{H}\alpha}_{\rm{esc,LyC}}$ is essentially the same but estimated assuming a transparent IGM line of sight.}

\begin{tabular}{p{3cm}p{3cm}}
\hline
Column name & Value \\ \hline

ID (in UVIT F154W catalog)                                                             &      13753                                                                                                                          \\
CANDELS ID                                               &      98192                                                                                                                          \\
MUSE ID                                                              &      899                                                                                                                          \\

RA, Dec (J2000)                                                                   & 53.15672, -27.79562                                                                                                            \\
$z$                                                                   & 1.097                                                                                                            \\
F154W (LyC) S/N                                                                   & 4.15                                                                                                            \\
$\text{H}\alpha^{1}$ line flux                                                                  & {$90.88\pm 3.18$}                                                                                                               \\
H$\beta$ line flux                                                               & {$28.53\pm10.65$} 
 \\

H$\gamma$ line flux                                                               & {$15.03\pm0.61$} 
 \\
 
{[O III]} $\lambda \lambda 4959, 5007$ line flux                                                          & {$56.83\pm 5.46$}                                                                                                 \\
{[O II] $\lambda3727$} line flux                                                           & {$77.49\pm0.97$}                                                                                            \\

$\text{EW}_{\text{H}\alpha,\text{rest}}$ (\text{\AA})  & $103.12\pm 5.72$ 
\\
{$\text{EW}_{\text{H}\beta,\text{rest}}$} (\text{\AA})  & {$16.81\pm 6.84$} 
\\
$\text{EW}_{\text{[O III]},\text{rest}}$ (\text{\AA})                                                        & $30.79\pm3.53$ \\
{Reduced $\chi^{2}$ of best-fit SED models}                                                        & {0.98 (G1), 0.78 (G2)} \\

{$\text{log}_{10}(\text{M}_{*}/\text{M}_{\odot})$}                                                           & {10.32$\pm$0.09 (G1)} \\ & {10.27$\pm$0.10 (G2)}                                                                                  \\
{$\text{SFR}_{\text{SED,10Myr}}$ ($\text{M}_{\odot}\ \text{yr}^{-1}$)}                                                          & {33.27$\pm$ 7.09 (G1)}, \\ & {10.37$\pm$7.54 (G2)}\\

$\text{SFR}_{\text{H}\alpha}$ ($\text{M}_{\odot}\ \text{yr}^{-1}$)
             & {$47.16\pm 8.71$}\\

{$\rm{E(B-V)}_{\text{neb, Balmer}}$}                                                               & {$0.55\pm 0.02$}\\
{$\rm{E(B-V)}_{\text{neb,SED}}$}                                                               & {0.5 (G1), 0.5 (G2)}\\
{$\rm{Z}_{\rm{SED}}$}                                                            & {0.006 (G1), 0.048 (G2)}                                                                                               \\
{$\text{log}_{10}(\xi^{\text{H}\alpha, \text{case B}}_{\rm{ion}}/\text{Hz\: erg}^{-1}$) } (G1)                                                            & {25.45$\pm$0.10} \\                    {$\text{log}_{10}(\xi^{\text{H}\alpha, \text{case A}}_{\rm{ion}}/\text{Hz\: erg}^{-1})$} {(G1)}  &     {25.84$\pm$0.10}                                         \\
$f^{\text{H}\alpha, \text{case B}}_{\rm{esc,LyC}}$, $f^{\text{H}\alpha, \text{case A}}_{\rm{esc,LyC}}$                                                            & {$0.12\pm0.03$} (case B) \\ & {$0.05 \pm 0.01$} (case A)                                                                     \\
$\tilde{f}^{\text{H}\alpha, \text{case B}}_{\rm{esc,LyC}}$,$\tilde{f}^{\text{H}\alpha, \text{case A}}_{\rm{esc,LyC}}$                                                      & {$0.08\pm0.02$} (case B) \\ &  {$0.03 \pm 0.01$} (case A)                                                                                                                           \\

{$f_{\rm{esc},\text{SED}}$}                                                         & {0.24 $\pm$ 0.12 (G1)}                            \\

\label{tab:galaxy10_prop}

\end{tabular}
\end{table}

\section{Data and Object Selection} 
\label{sec:data_sample}
We use HST (HLF, \cite{Whitaker19}), JWST (JADES, \cite{Rieke23}) and UVIT (AUDFs-GOODS South Survey, Saha et al. 2024, under review) images  of the GOODS-South field. The PSF FWHM=1.6" and the 3$\sigma$ depth is 27.2 mag in the UVIT F154W band.
\par We use the CLEAR grism survey catalogs \citep{Simons23} to query galaxies in the GOODS-South field between redshifts 1 and 1.5, having at least the H$\beta$, [O III] and H$\alpha$ emission lines with a signal-to-noise ratio (S/N) greater than three and that lie within the coverage of the UVIT GOODS-South images. The lower limit of the redshift criterion is chosen based on the red cut-off of the UVIT F154W band ($\lambda_{\text{eff}}=1541 \text{\AA}, \lambda_{\rm{max}}=1800 \text{\AA}$). Thus the F154W band samples LyC photons at redshifts greater than 0.97. The upper limit of the redshift criterion is chosen to ensure that the H$\alpha$ line,  
falls within that region of the G141 grism coverage, that has good sensitivity. We find a total of 95 such objects. 
\par We then visually examine high-resolution HST band and/or JWST band (when available) images and select those objects that show signs of interaction with a companion galaxy. We use visual signatures like the identification of double nuclei, stellar bridges, tidal tails, and plumes \citep{Kartaltepe15,Martin22}. We identify 24 merger candidates at this stage. Next, keeping in mind the resolution of the UVIT F154W band, we discard objects with potential contaminants within an aperture of radius $1.6"$ from their center. Foreground contaminants, are differentiated from objects that are interacting with the system based on the absence of visual signatures of interaction as well as an examination of the spectroscopic data when available. To aid the understanding of our selection criterion in this step, we highlight examples of discarded candidates in Appendix \ref{sec:eg_discard}. This step is important given the susceptibility of foreground contamination in LyC leaker searches using instruments with coarser resolution \citep{Vanzella10}. We are left with 13 objects. \par In the last step, we proceed to query the AstroSat UV-Deep Field GOODS-South (AUDFs) F154W catalog (Saha et al. 2024, under review) at the positions of the remaining 13 objects and select those for which there lies a detected object within $\sim 1.6"$. The remaining sample after this step consists of just a single object, {AUDFs\_F13753} (hereafter AF13753). \par
\begin{figure}[ht!]
\nolinenumbers
\hspace{-0.4cm}
\includegraphics[width=1\columnwidth]{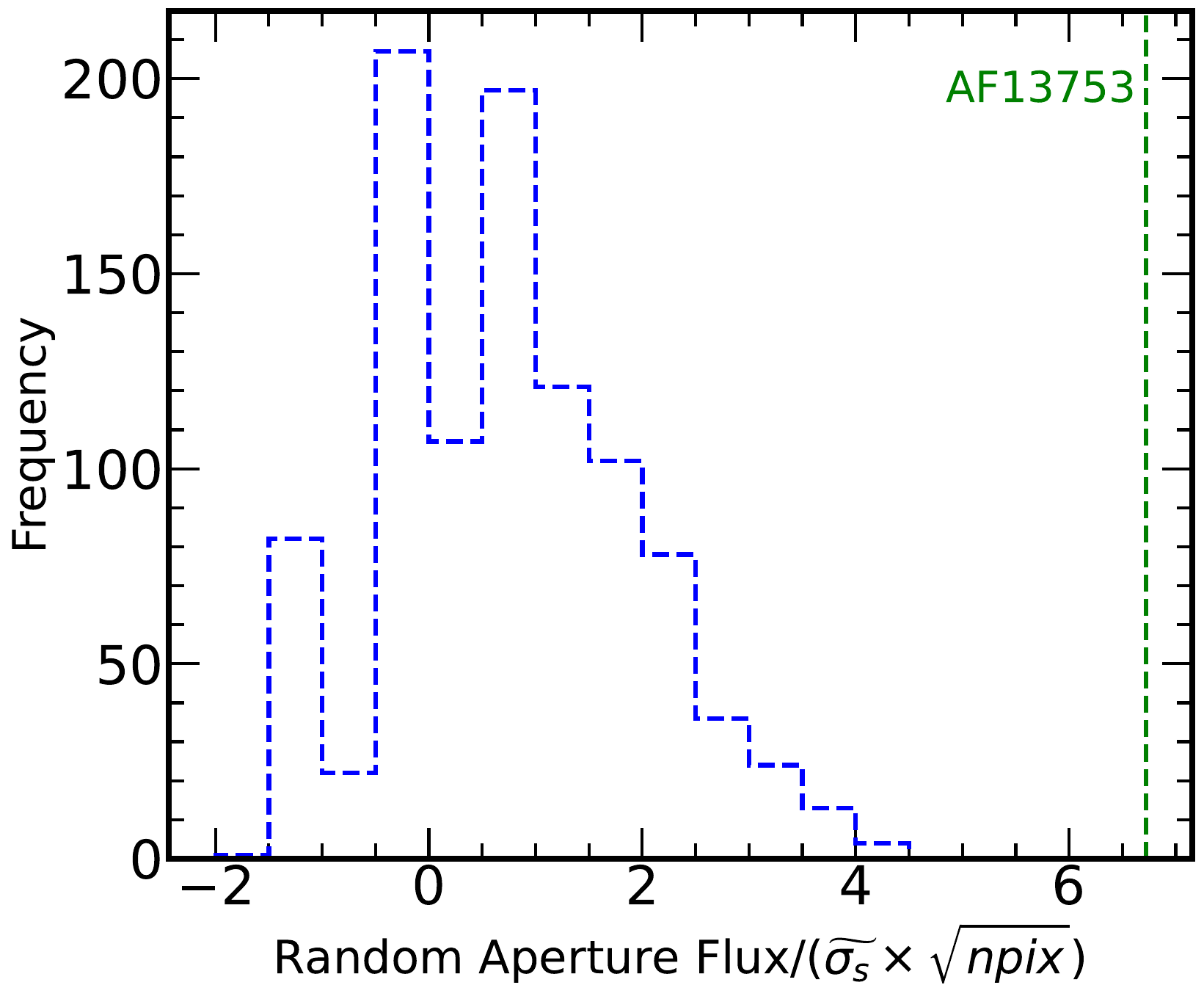}
\caption{Normalized F154W flux of AF13753 (6.73), indicated with the green dashed line with respect to the distribution of fluxes (blue dashed histogram) within 1000 apertures placed across empty regions of the UVIT F154W GOODS-South image. The flux units are normalized to the sky background rms ($\widetilde{\sigma_s}$) and $npix$ denotes the number of pixels within the aperture. Note that the normalized flux of AF13753 is estimated within an aperture of diameter PSF FWHM (1.6") in this particular experiment.} 
\label{fig:galaxy10_sig}
\end{figure}

 At the location of AF13753, the 5$\sigma$ depth in the UVIT F154W band, using an aperture of radius PSF FWHM ($\sim$ 1.6") is 26.64 mag. AF13753, with a magnitude of 26.36 estimated using an aperture of radius PSF FWHM, is detected at 6.15 significance above the local background rms. The S/N ($\sim 4$) of this object is estimated using the S/N formula in Saha et al. 2024, where the Poisson shot noise of the source is also considered. In Saha et al. 2024 we also try to quantify the noise in the UVIT images across the entire field by carrying out an empty aperture analysis, where we place apertures of diameter FWHM across 1000 empty regions of the field. Using the flux of the AF13753 within an aperture of diameter FWHM, we find its significance to be 6.73 in this experiment. Figure \ref{fig:galaxy10_sig} highlights the normalized flux of AF13753 with respect to the distribution of the same within the empty apertures.

The system is thus selected based on the available spectroscopic information confirming its redshift ($z=1.097$) from both the CLEAR \citep{Simons23} and MUSE HUDF surveys \citep{Bacon23}, visual inspection of its morphology and potential foreground contamination, and its high S/N in the F154W band.

\subsection{Spectroscopic data and measurements}
We obtain the necessary spectroscopic data products for AF13753 from the MUSE Hubble Ultra Deep Field survey \citep{Bacon23} and the HST-CLEAR grism survey \citep{Simons23}. This gives us spectral coverage in the rest-frame of the object from $2241-8107 \text{\AA}$. We detect multiple emission and absorption lines enabling us to confirm its redshift. We highlight all the emission and absorption lines we are able to identify in the MUSE 1D spectra of AF13753 in Figures \ref{fig:galaxy10_MUSE_1D} and \ref{fig:galaxy10_OII_v2} in Appendix \ref{sec:muse1d}. We confirm the redshift of 1.097 and we do not report any lines that are inconsistent with this redshift.
\par For the 3D-HST CLEAR grism data, we carry out the emission line fitting and generate H$\alpha$ and [O III] emission line maps (Figure \ref{fig:galaxy10_images}) using the software \texttt{grizli}\footnote{\href{https://github.com/gbrammer/grizli/}{https://github.com/gbrammer/grizli/}} \citep{Brammer19}. The emission line fits and the redshift estimation in \texttt{grizli} are carried out simultaneously by fitting a nonnegative linear combination of a basis set of the continuum Flexible Stellar Population Synthesis model templates by \cite{Conroy09} over which emission lines and emission line complexes are included. The final best-fit model is obtained from a $\chi^2$ minimization over the grid of trial redshifts. At a given trial redshift, \texttt{grizli} uses the multiwavelength HST photometry. The spectra are scaled to the photometry. The imaging data is also used to ascertain the spatial morphology of the source and the redshifted templates are forward modeled into 2D grism spectral beams. Single component fits are used to fit the [S II] $\lambda \lambda6718$ and 6732, and the [O III] $\lambda \lambda 4960$ and 5007 doublets. For further details, we refer the reader to \cite{Simons21,Simons23}. The [O III] doublet is not resolved in the grism data. We highlight the G102 and G141 grism spectra and the {best-fit line templates} in the left panel of Figure \ref{fig:galaxy10_grism1d}. \\ 
Using the theoretical line ratio of 2.98 for the [O III] doublet \citep{Storey00}, we separately estimate the flux of the [O III] $\lambda$5007 line whenever required.
\par  Although the HST grism spectra offers a superior spatial resolution compared to the MUSE IFU spectra, the latter is significantly deeper. The $3 \sigma$ line flux limit for the MUSE spectra is $\sim 3.1 \times 10^{-19} \text{erg}\: \text{s}^{-1}\text{cm}^{-2}$, \citep{Bacon23} while the same for the G141 grism spectra is $\sim 2.1 \times 10^{-17} \text{erg}\: \text{s}^{-1}\text{cm}^{-2}$, \cite{Momcheva16}.

\section{Analysis and Results}
\subsection{Structure and Morphology} 
\label{sec:morph} 
\begin{figure*}[ht!]
\nolinenumbers
\includegraphics[width=1\textwidth]{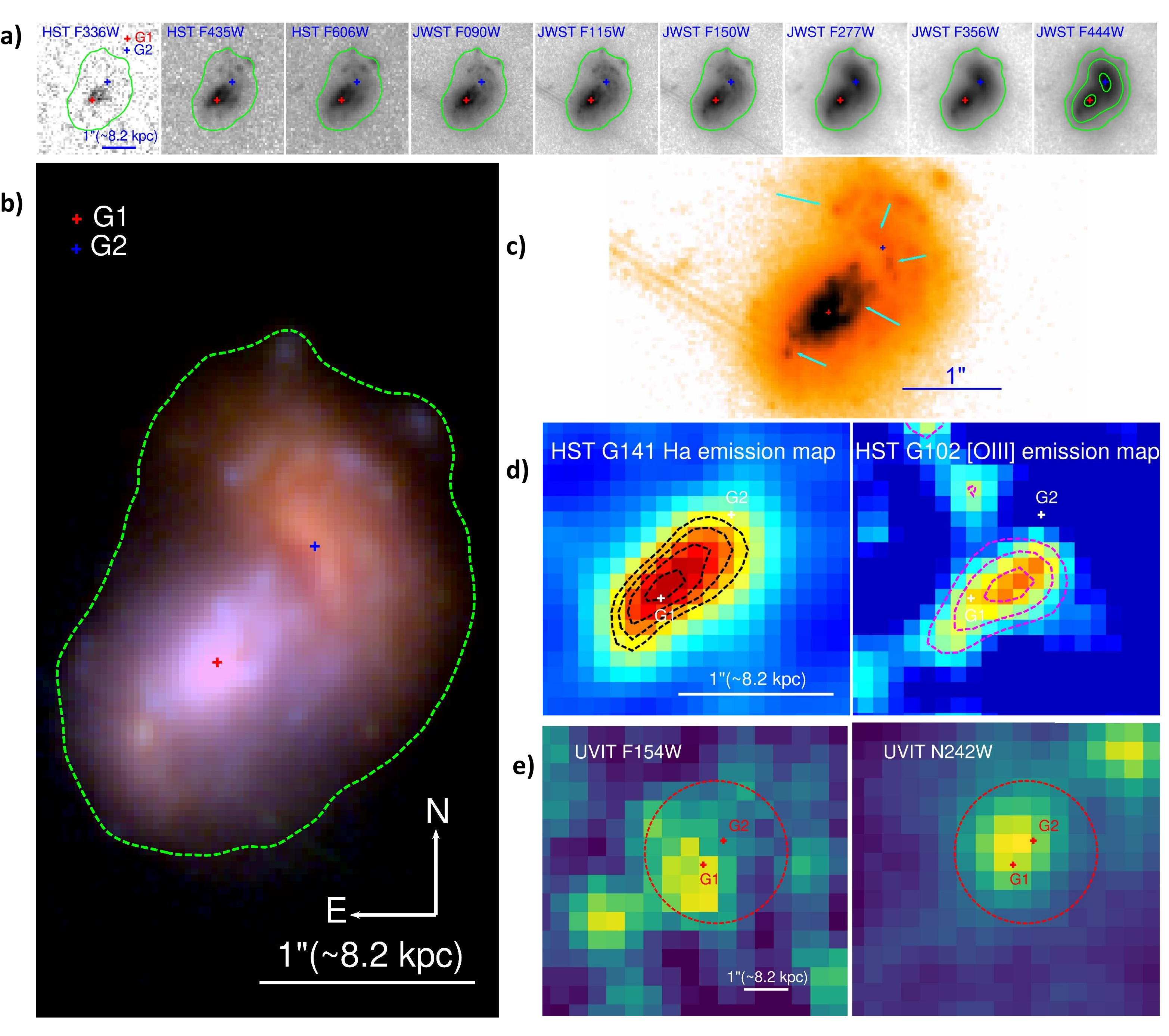}
\caption{Multiband images of AF13753. North is up and East is to the left in all the images. The positions of the two galaxies are marked in each image by crosses. Panel (a) displays the galaxies in the optical and infrared bands, covering the restframe FUV to NIR emission.  
The outermost  $\sim 12 \sigma$ contour (in green) in the F444W band is plotted onto the bluer band images to highlight the extent of the interacting system. Panel (b): RGB color image constructed using the JWST NIRCam bands - F277W (red), F115W(green) and F090W (blue). Panel c) highlights some of the disturbed features in the central regions of the G1-G2 system (cyan arrows) in the rest-frame optical band (JWST F115W). Panel (d) displays the HST G141 H$\alpha$ and G102 [O III] emission line maps. The H$\alpha$ and [O III] contours are marked in black and magenta respectively. Panel (e): UVIT imaging of AF13753. The radius of the red dashed aperture is equal to the F154W PSF-FWHM ($1.6"$). The UVIT images have been smoothed with a Gaussian kernel of radius 2 pixels. The HST and JWST band images (Panels a and b) are displayed in log-scale.} 
\label{fig:galaxy10_images}
\end{figure*}

In Figure \ref{fig:galaxy10_images}, we display the multi-wavelength imaging data for the system AF13753. The visual morphology suggests that it consists of two distinct galaxies in the rest-frame near-infrared (NIR) bands observed JWST/NIRcam. These two galaxies are separated by $\sim 5.5$~kpc ($\sim 0.7"$ on sky). We label the southern galaxy G1 and the northern galaxy G2. Their positions are determined with respect to the JWST NIRCam detection band. We run Source Extractor \citep{Bertin96} to distinguish the two galaxies and identify their on-sky positions which are marked in all the panels of Figure \ref{fig:galaxy10_images}. 

{Visual inspection of the rest-frame UV/optical and IR images taken by HST and JWST (see Panel a in Figure~\ref{fig:galaxy10_images}) show wide-spread clumpy structures. The RGB color image (Panel b) constructed using the JWST NIRCam filters reveals that the galaxy G1 is bluer compared to G2. Both G1 and G2 appear to be surrounded by an envelope of diffuse stellar emission. This is clearer in the rest-frame optical ($\lambda \sim 5542\:\text{\AA}$) band probed by the JWST NIRCam F115W filter (Panel c). The stellar envelope is elongated along the NW-SE direction which is meaningful given that it contains both the galaxies. However, on closer inspection, there is an asymmetry along the NE-SW direction with respect to the center of G1. In addition, there are filamentary clumpy structures in the central regions of G1 and G2, marked by the cyan arrows in Panel c). These clumpy structures are significantly elongated with projected sizes in the range of $\sim 2 - 4$~kpc. The asymmetry of the stellar envelope and clumpy structures have been used to infer ongoing mergers of galaxies in the high-redshift universe \citep{Elmegreenetal2009, Ribeiroetal2017, Martin22}. The absence of a tidal tail in the interacting galaxy-pair could indicate that this might be the case of a retrograde merger between two disk galaxies in which case a prominent tidal tail may not develop at all. Nevertheless, based on visual classification schemes \citep{Veilleuxetal2002,Larsonetal2016}, the projected separation of 5.5 kpc and theoretical support from gas dynamical simulations \citep{MihosHernquist1996}, the interacting galaxy pair G1-G2 can be classified as being in an M2 interaction stage, i.e., in a close binary pre-coalescence stage \citep{Veilleuxetal2002, Hungetal2015}. The clumpy nature of the galaxy pair in the rest-frame UV/optical filters suggest a strong starburst - possibly triggered by the ongoing merger.}
\par The starburst nature is also evident from the presence of strong emission lines such as H$\alpha$, [OIII], [OII] observed in this galaxy pair. The H$\alpha$ and [O III] emission line maps of the system (Panel d) indicates G1 undergoing concentrated star-formation activity. The LyC emission from this interacting system is traced by the F154W band (corresponding to mean $\lambda_{\rm{rest}}=734$\AA) and interestingly, the LyC emission seems concentrated on G1 (Panel e). The rest-frame far-UV emission of this system is probed by the N242W band of UVIT.

\subsection{Ionized gas kinematics} \label{sec:kin}

\begin{figure*}[ht!]
\nolinenumbers
\centering
\includegraphics[width=2.25\columnwidth]{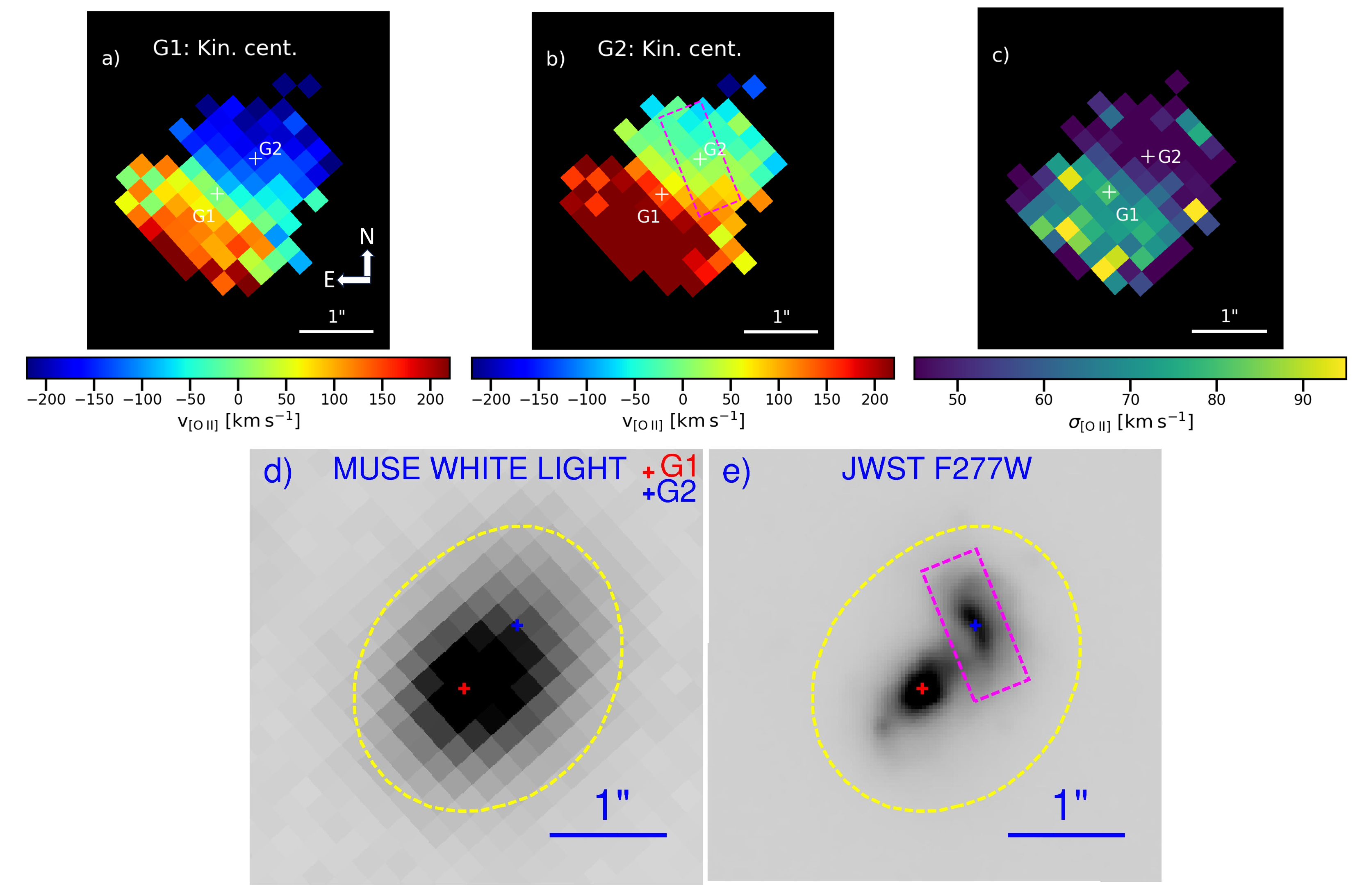}
\caption{{[O II] velocity (Panels a and b) and dispersion  (Panel c) maps of AF13753 constructed using the MUSE data cube. The location of G1 and G2 are indicated by the white crosses in the kinematic maps. G1 and G2 are chosen as the kinematic centers of the velocity maps in Panel a) and b) respectively. The MUSE white light image of the system is shown in Panel d), with the yellow dashed curve indicating the contour of $7 \sigma$ in this image. The contour in the MUSE white light image and the dashed purple box in Panel b) is overplotted on the JWST F277W image in Panel e). North is up and East is to the left in all the panels as in Figure \ref{fig:galaxy10_images}. }}  
\label{fig:galaxy10_muse1d}
\end{figure*}

{In Panels a), b) and c) of Figure \ref{fig:galaxy10_muse1d}, we highlight the ionized gas kinematics of the system that is traced by the [O II] doublet. For this purpose, we extract and fit the [O II] $\lambda \lambda 3727, 3729$ lines for each spaxel of the system in the MUSE data cube. We fit each [O II] doublet using two Gaussians, keeping the relative redshifts between the $\lambda \lambda 3727, 2729$ lines fixed. For our final fits, we include only those spaxels for which the [O II] line has a S/N greater than 3. In addition to the S/N cut, we manually check and remove spaxels with bad fits. The errors in the fitted velocity (dispersion) are of the order of $\sim 12\: \rm{km}\: \rm{s}^{-1}$ ($\sim 3\: \rm{km}\: \rm{s}^{-1}$) in majority of the map area. Towards the boundaries of the map the errors are of the order of $\sim 40\: \rm{km}\: \rm{s}^{-1}$ ($\sim 15\: \rm{km}\: \rm{s}^{-1}$). We adopt the line-spread function from \cite{Bacon23} when determining the velocity dispersion.}

\par {The spatial resolution of the MUSE observations (FWHM $\sim 0.5"$ at 8000 \AA, \cite{Bacon23}) means that the G1-G2 system (Panel e) remains unresolved (Panel d) in the MUSE data. Thus, given the limited spatial resolution of MUSE, and the relative faintness of G2 in the rest-frame ultraviolet/optical bands, the following question is pertinent: Can the observed features in the kinematic maps be simply explained by a normal rotating disk associated with G1 after considering the MUSE PSF smearing? The kinematic distinction between the orbital motion of mergers and rotating disks is an outstanding problem, especially when using seeing-limited data \citep{Simons19}. Techniques to discern asymmetries in the velocity and dispersion fields often require high signal-to-noise and spatially resolved data in order to decompose the various modes of the kinematic fields \citep{Krajnovic06,Shapiro08,Hungetal2015}. For simplicity, we compare our kinematic maps to the case of a normal rotating disk, leaving room for the possibility of a more complex kinematic structure that can be associated with the observed velocity and dispersion map.}

\par {To aid the visualisation, we adopt two choices of the kinematic center (Panels a and b), in the velocity map. The first choice is the center of G1 (Panel a). This is nearly coincident with the centroid of the MUSE continuum corresponding to the [O II] emission (difference $\sim$ 1 spaxel). At a first glance, the kinematic map corresponding to this center (Panel a) suggests a normal rotationally supported system, {akin to what has been observed in some of the M2-type interacting galaxy pairs e.g., IRASF12043-3140 and IRASF21330-3846 \citep{Hungetal2015} }. However, a close inspection of the velocity map reveals a significant deviation, along the southern side of G2, from the expected ordered velocity field of a normal rotating disk characterized by the isovelocity contour features well-known as the spider diagram (e.g., Figure 1 of \cite{Kruit78} and \cite{Shapiro08}). In other words, the asymmetry along the NE-SW direction in the projected velocity map is not consistent with the velocity map of a normal rotating disk galaxy. The effect of this asymmetry is better reflected  in Panel b, where we 'force' the kinematic center to be the location of G2. The dashed purple box centered at G2 is roughly along the photometric major axis of G2 (obtained in the GALFIT modeling in Section \ref{sec:sed}). Within the box marking G2, there is a hint of rotation, albeit lopsided. The asymmetry in the radial velocity along the NE-SW direction (which is roughly along the G2-major axis) is persistent, with the SW side appearing to move $\sim 50\: \rm{km}\: \rm{s}^{-1}$ faster than the NE side.} 
\par {We find evidence of a gradient in the radial velocity dispersion field when going from G1 to G2, in Panel c). This is atypical for an ideal relaxed rotating disk, where one would expect the dispersion to peak near the kinematic center and decrease in a near symmetric fashion on either side from this peak, as is seen in several disky galaxies at $z\sim 0.9 - 2.7$ \citep{Wisnioski15,Harrison17}. The relatively higher velocity dispersion on the side of G1 could be triggered by the star-forming clumps themselves \citep{ElmegreenBD2005}. Given these anomalies, we conclude that the observed gas kinematic features are unlikely to be explained by a simple rotating disk associated with G1. The observed complex kinematics are possibly induced by the merger, disk perturbations or the effects of both \citep{Hungetal2015}.}

\subsection{Photometry and SED modeling} 
\label{sec:sed}
In this section, we perform a photometric analysis of G1 and G2 with an aim of determining photometric redshifts and of inferring physical properties of their stellar populations through SED modeling. Here, we use the following photometric filters: UVIT F154W; HST F275W, F336W, F336W, F606W, F435W, F606W, F775W, F814W, F850LP, F105W, F125W, F140W, F160W; JWST F115W, F150W, F200W, F277W, F335M, F356W, F410M, and F444W. We attribute the flux in the UVIT F154W band, that traces the LyC emission of the system to only the galaxy G1. This choice is motivated by the morphology of the H$\alpha$ and [O III] emission and the fact that G1 dominates the rest-frame FUV emission of the system (Figure \ref{fig:galaxy10_images}). For the HST and JWST photometry, we first PSF-match the images in these bands to the HST F160W band image. The complex morphology of the system in addition to the smoothing introduced by the PSF matching means there could be a non-negligible contamination from the light of G1 while performing the photometry of G2 in the PSF-matched images and vice versa. This, in turn, could potentially influence our SED fitting. To minimize this effect, we carry out a simultaneous fitting of the light profiles of the entire system in the PSF-matched bands using the 2D fitting software GALFIT \citep{Peng02}. Owing to the limited angular resolution, we opt for a simple fitting procedure using two S\'ersic profiles representing G1 and G2. Due to the inhomogeneous nature of G2 in the bluer HST bands, we keep the position angle and the centroid of the S\'ersic profile of G2 in these bands fixed based on the fitted values of these parameters in the JWST NIRCam F444W band. However, in the bluest HST F275W and F336W bands we are unable to obtain a suitable fit for G2. We choose to carry out a single S\'ersic fit for G1 in these bands and exclude these two bands for the SED fitting of G2. The resulting GALFIT model fluxes for G1 and G2 are employed in the subsequent SED fitting.  
\par Since the grism emission line maps do not indicate H$\alpha$ and [O III] emission from G2, we use the model fluxes to evaluate the photometric redshift of G2. We use \texttt{EAZYpy} \citep{Brammer08,Brammer21} for this purpose. Given a set of templates, \texttt{EAZYpy} uses nonnegative linear combinations of these templates to fit the photometric data and returns the probability density function for the photometric redshift. We use the PEGASE2.0 template set for the photometric redshift estimation of G2, since it includes models with varying star-formation histories, ages and large dust-content that represent dusty star-forming objects. We obtain a photometric redshift of $1.25^{+0.22}_{-0.18}$ for G2. This value is consistent within the error with the spectroscopic redshift of $\sim 1.1$ which we determine from the MUSE and CLEAR spectra. 
\par For the SED modeling, we adopt the latest version of CIGALE \citep{Boquien19,Yang22}. Our CIGALE run follows \citet{Saha20}, with the addition of now performing the fitting over a range of gas phase metallicities. Briefly, we use the BC03 stellar population, a Salpeter IMF, and an exponentially declining SFH with late bursts. We use the \textit{dust\_att\_modified\_starburst} module to model the dust attenuation. We allow the {nebular} ($\rm{E(B-V)}_{\rm{neb}}$) reddening values to vary for both G1 and G2. We provide CIGALE with the following range of values for the $\rm{E(B-V)}_{\rm{neb}}$; $[0.1,0.3,0.5,0.8,1,1.5]$. We however, keep the conversion factor from the nebular to the stellar reddening fixed to 0.44 \citep{Calzetti00}.  

\begin{figure*}[ht!]
\nolinenumbers
\centering
\includegraphics[width=2\columnwidth]{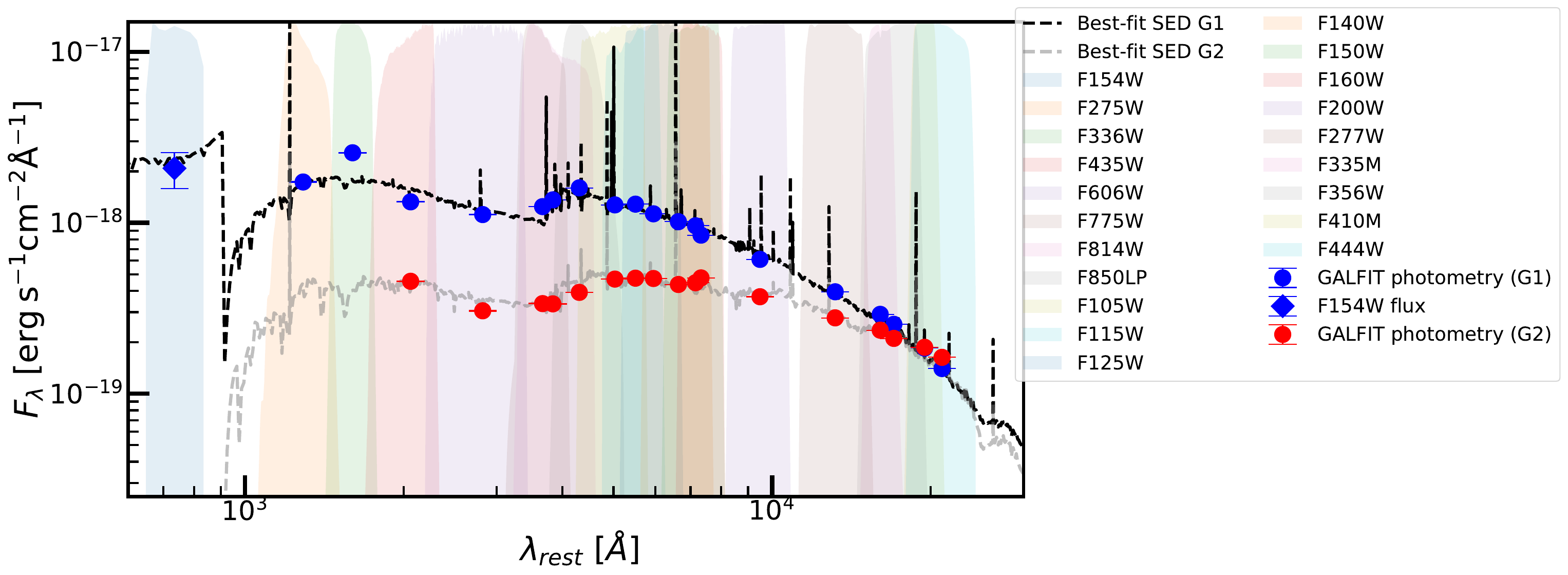}
\caption{The best fit SED models of G1 (black) and G2 (grey). The F154W flux is denoted by the blue diamond. Both the x and y axes are in logarithmic scale. In the background, the colored curves denote the shape of broadband filters adjusted for the rest-frame wavelength.} 
\label{fig:galaxy10_g1_g2_sed}
\end{figure*}

For simplicity, we allow for variable escape fraction only for G1 and adopt an escape fraction of zero for G2 in the fitting. 
The best-fit SED models are shown in Figure \ref{fig:galaxy10_g1_g2_sed} and some of its parameters are included in Table \ref{tab:galaxy10_prop}. \par The SED-derived stellar masses of G1 and G2 are similar which implies that this is a major merger with mass ratio, {$\text{M}_{*,\text{G1}}/\text{M}_{*,\text{G2}}=1.13 \pm 0.37$}. To the best of our knowledge, this work reports the first detection of LyC emission specifically from a major-merger. The best fit SEDs support G1 having a lower metallicity and higher star formation than G2 (Table \ref{tab:galaxy10_prop}). The best-fit SED of G2 suggests that it is dust rich like G1 ($\rm{E(B-V)}=0.5$) and has a super-solar metallicity with weak emission lines (Figure \ref{fig:galaxy10_g1_g2_sed}).

\subsection{Dust attenuation} \label{sec:dust} 
In addition to the reddening value from the best-fit SED, we estimate the nebular reddening $\rm{E(B-V)}_{\text{neb}}$ using the observed H$\alpha$ and H\pmb{$\gamma$} line fluxes. In the grism spectra, the H$\beta$ line is detected with an S/N=2.68. However the H$\gamma$ line is detected in the MUSE spectra with an S/N=24.72. Thus we use the H$\alpha$ to H$\gamma$ ratio to estimate the reddening. Following \cite{Sobral12}, we correct the H$\alpha$ line in the grism spectra for [NII] contamination using our measured $\text{H}\alpha+\text{[NII]}$ equivalent width obtained using \texttt{grizli}. The relation is given by the following polynomial equation:
\begin{equation}
    y=-0.924+4.802x-8.892x^2+6.701x^3-2.27x^4+0.279x^5,
\end{equation}

\noindent where $y=\rm{log}(\text{F}_{[\rm{NII}]}/\text{F}_{\rm{H}\alpha})$ and $x=\rm{log}(\rm{EW}_{\rm{rest}}(\rm{H}\alpha+[\rm{NII}])/\text{\AA})$.
We find $\text{F}_{[\rm{NII}]}/\text{F}_{\rm{H}\alpha} \sim 0.09$. We obtain an $\text{H}\alpha/\text{H}\gamma$ ratio of $13.9\pm0.04$. Assuming a case B recombination and temperature $T=10^4 \text{K}$, we estimate {$\rm{E(B-V)}_{\text{neb}}$} $=0.55 \pm 0.06$. \footnote{For comparison, using the H$\alpha$ and H{$\beta$} ratio, $\rm{E(B-V)}_{\text{neb}}$ $=0.67 \pm 0.42$.} This is in good agreement with the best fit $\rm{E(B-V)}_{\text{neb}}$  from the SED fitting. 
\par We employ the above obtained value of $\rm{E(B-V)}_{\text{neb}}$ in conjunction with the nebular extinction curve of \cite{Reddy20} to estimate the dust-corrected luminosity of the H$\alpha$ line which is subsequently used to estimate the $\text{SFR}_{\text{H}\alpha}$ \citep{Kennicutt98}. Based on the emission line maps and the rest-frame UV emission, we assume that the H$\alpha$ SFR ($47.16 \pm 1.65 \text{M}_{\odot}\ \text{yr}^{-1}$) predominantly traces the star formation in G1. We compare the H$\alpha$ SFR to the main sequence relation of \cite{Whitaker14} that uses H$\alpha$ as the SFR indicator after accounting for differences in the IMF calibration. To be consistent with the H$\alpha$ SFR measurements of \cite{Whitaker14}, we do not account for the dust correction of H$\alpha$ in this comparison. We find that G1 lies above the main sequence indicating that it is undergoing a starburst. 
We discuss the implications of dust attenuation to the escape of ionizing photons in Section \ref{sec:geometry}.

\subsection{AGN diagnostics} \label{sec:agn_diag}
The Chandra 7 Ms survey \citep{Luo17} contains a detection with an S/N$\sim3$ in the soft X-ray band, i.e., in the 0.5-2 keV band (restframe $\sim$ 1.05-4.2 keV) associated with AF13753. It is not detected in the harder, 2-8 keV band (restframe $\sim$ 4.2-16.8 keV). {The aperture-corrected source counts reported in the Chandra catalog in the soft X-ray band is $12.5^{+5.5}_{-4.3}$. The X-ray luminosity corresponds to $\rm{log}(\rm{L}_{\rm{X,int}}/\rm{erg}\: \rm{s}^{-1})\approx 41.34$}. The source in the Chandra catalog is classified as a normal galaxy and not an AGN. {\cite{Luo17} classify a source as an AGN if it satisfies any one of the six criteria they outline in their work. Five of these criteria are reliant on the detected X-ray fluxes. Luminous X-rays sources are classified as AGNs if they have an $\rm{log}(\rm{L}_{\rm{X,int}}/\rm{erg}\: \rm{s}^{-1})>42.48$. This criteria alone however, rules out low-luminosity or dust obscured AGN. Sources with an effective power-law photon index ($\Gamma_{\rm{eff}}\leq 1$) are selected as hard X-ray sources. Since AF13753 is not detected in the hard X-ray band, \cite{Luo17} derive the probabilistic best guess estimate of $\Gamma_{\rm{eff}}$ for such sources, i.e.,  sources that are detected in only one X-ray band. For AF13753 this estimate of $\Gamma_{\rm{eff}}$ is 2.28. The remaining four criteria, any of which if satisfied are used to classify a source as an AGN are the following: (1) if the X-ray-to-optical flux ratio $\rm{log}(f_{\rm{X}}/f_{\rm{R}})>-1$, (2) if the object is spectroscopically classified as an AGN, (3) if the X-ray-to-radio luminosity ratio $\rm{L}_{\rm{X,int}}/\rm{L}_{1.4\rm{Ghz}}\geq 2.4 \times 10^{18}$ (not applicable for AF13753), (4) if the X-ray-to-NIR flux ratio, $\rm{log}(f_{\rm{X}}/f_{\rm{Ks}})>-1.2$. $f_{\rm{R}}$ and $f_{\rm{Ks}}$ correspond to the R and Ks band fluxes respectively. {For AF13753, the values of ${\rm{log}(f_{\rm{X}}/f_{\rm{R}})}$ and ${\rm{log}(f_{\rm{X}}/f_{\rm{Ks}})}$ are $\sim$ -2.39 and -2.85 respectively.}}
\par {The soft-X ray emission of AF13753 may arise from galactic winds driven by either starbursts or nuclear activity, both of which can be triggered by the merger (e.g., \cite{Kukreti22}). We compare the soft X-ray luminosity of AF13753 to that of a sample of nine interacting spiral galaxies in the local universe \citep{Brassington06} probed at 0.3-6.0 keV. We find comparable values to that of AF13753, with $\rm{log}(\overline{\rm{L}}_{\rm{X,int}}/\rm{erg}\: \rm{s}^{-1}) \sim 41.41$ for this sample, where $\overline{\rm{L}}_{\rm{X,int}}$ is the mean X-ray luminosity of the sample.}

\par
\begin{figure*}[ht!]
\nolinenumbers
\centering
\includegraphics[width=2.15\columnwidth]{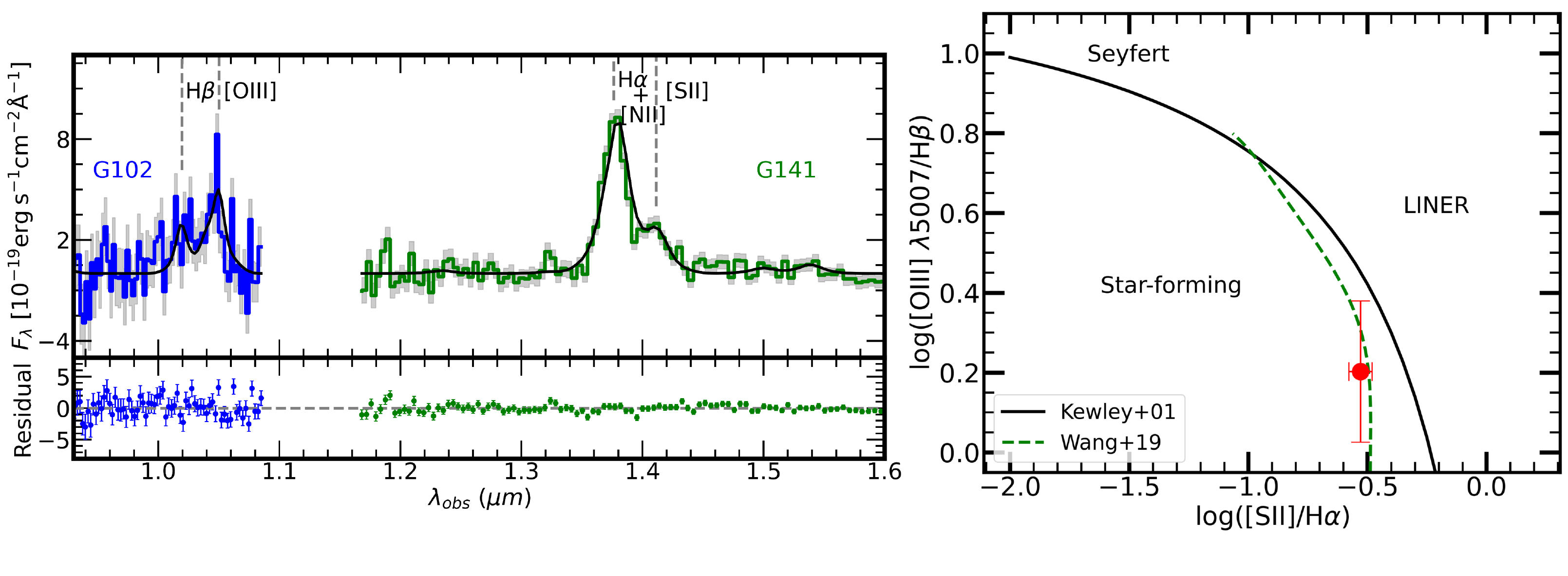}
\caption{Left panel: The extracted grism 1D spectra of AF13753. The G102 grism spectrum is shown in blue, and the G141 grism spectrum in green. The spectra are continuum subtracted. The grey vertical lines indicate the identified emission lines. The grey-shaded curve indicates the error spectrum. {The black curve denotes the best-fit lines obtained using \texttt{grizli}}. The bottom panel denotes the corresponding residuals. Right panel: Location of AF13753 on the BPT diagnostic diagram of \cite{Kewley01} using the line fluxes from the grism analysis. {The green curve denotes the fit of \cite{Wang19} used to quantify the [S II] deficiency (see Section \ref{sec:geometry}).}} 
\label{fig:galaxy10_grism1d}
\end{figure*}

 In the right panel of Figure \ref{fig:galaxy10_grism1d}, we locate the position of AF13753 in the $\rm{log}([\rm{S\:II}]/\rm{H}\alpha)$ vs $\rm{log}([\rm{O\:III}]/\rm{H}\beta)$ BPT diagnostic diagram \citep{Baldwin81} adapted from \cite{Kewley01}. AF13753 occupies the star-forming region in the plot. The line fluxes used are measured from the grism spectra. {We note that since our estimate of the ratio of the H$ \alpha$ to [N II] is only indirectly inferred 
(Section \ref{sec:dust}), we do not include the diagnostic with $[\rm{N\:II}]/\rm{H}\alpha$.} 

\subsection{Ionizing photon production and escape fraction}
We utilize the H$\alpha$ and the best-fit SED parameters to estimate the ionizing photon production efficiency ($\xi_{\rm{ion}}$) and  $f_{\rm{esc,LyC}}$ for the galaxy G1.  $\xi_{\rm{ion}}$ quantifies the LyC photon production rate per unit UV luminosity density in the galaxy and is an important ingredient for reionization modeling \citep{Robertson22}.
Assuming case B recombination, temperature $T=10^4 \text{K}$ and $n_e=100 \: \text{cm}^{-3}$, we first estimate the intrinsic rate of ionizing photons used in the recombination process \citep{Ferland06} which we define as the ones that do not escape the galaxy:

\begin{equation} 
\label{eq:N_nonesc}
    \dot{N}^{non-esc}_{\rm{LyC}, \text{case B}} [\:\text{s}^{-1}]=7.28 \times 10^{11} L_{int,\rm{H}\alpha} [\text{erg}\: \text{s}^{-1}] .
\end{equation}

\noindent In the above equation, $L_{int,\rm{H}\alpha}$ denotes the intrinsic H$\alpha$ line luminosity. Using $\dot{N}^{non-esc}_{\rm{LyC}, \text{case B}}$ and our best-fit SED, it is possible to estimate $\xi_{\rm{ion}}$ as:

\begin{equation} \label{eq:zi_ion}
    \xi_{\rm{ion}} [\text{Hz\: erg}^{-1}]=\frac{\dot{N}^{int}_{\rm{LyC}}}{L_{1500} },
\end{equation}

\noindent where $\dot{N}^{int}_{\rm{LyC}}$ denotes the total intrinsic LyC photon production rate and $L_{1500}$ is the luminosity density at rest-frame $1500\: \text{\AA}$ in units of $\text{erg}\  \text{s}^{-1} \text{Hz}^{-1}$. We estimate $L_{1500}$ using the intrinsic stellar unattenuated model component of the best-fit CIGALE SED. For our estimate of $\xi_{\rm{ion}}$ we replace $\dot{N}^{int}_{\rm{LyC}}$ with $\dot{N}^{non-esc}_{\rm{LyC}, \text{case B}}$ \citep[see][]{Simmonds23}; essentially in doing so we assume the escape fraction of ionizing photons is negligible and denote this quantity as $\xi^{\text{H}\alpha, \text{case B}}_{\rm{ion}}$ (see Table \ref{tab:galaxy10_prop}). Accounting for the escaping ionzing photons will increase the value of $\xi_{\rm{ion}}$. Since we detect LyC emission, the optically thick assumption of case B recombination may not be suitable. Therefore, in Table \ref{tab:galaxy10_prop}, we also provide estimates of $\xi^{\text{H}\alpha}_{\rm{ion}}$ assuming a case A recombination scenario.
\par We adopt two approaches to estimating the escape fraction \citep{Saha20}, the first one is based on the derived $\dot{N}^{non-esc}_{\rm{LyC}, \text{case B}}$ from the H$\alpha$ line and the second uses the best-fit SED (Section \ref{sec:sed}). The observed flux in the UVIT F154W band corresponds to the LyC flux at $\lambda_{\rm{rest,mean}}\sim 740 \text{\AA}$. The entire rest-frame wavelength range probed by the F154W band corresponds to 650-858 \text{\AA}.  Following \cite{Dhiwar24}, we convert the observed F154W flux into the escaping LyC photon rate (assumed to be from G1) into the IGM, which we denote as $\dot{N}^{esc}_{\rm{LyC}}$. The formula is given by, 

\begin{equation}\label{eq:NescLyC} 
    \dot{N}^{esc}_{\rm{LyC}} = \frac{L_{\rm{F154W}} \times \lambda^{\rm{F154W}}_{\rm{rest}}}{hc/\lambda^{\rm{F154W}}_{\rm{rest}}} \times \rm{exp}[{\tau_{\rm{IGM}}}] ,
\end{equation}

where $\lambda^{\rm{F154W}}_{\rm{rest}} = 1541\text{\AA}/(1+z)$ for the F154W filter, $L_{\rm{F154W}} \times \lambda^{\rm{F154W}}_{\rm{rest}}$ denotes the rest-frame luminosity in units of $\text{erg}\:\text{s}^{-1}$ computed using the observed flux in the F154W filter and the cosmological luminosity distance. 
Since the definition of ${\dot{N}^{esc}_{\rm{LyC}}}$ accounts only for the ionizing photons falling within the observed F154W band, this implies that the quantity as defined is a lower limit to the escaping ionizing photon rate. Estimating $\dot{N}^{esc}_{\rm{LyC}}$, additionally requires correcting for absorption by the IGM, which in the above formula (Equation \ref{eq:NescLyC}) is denoted by the $\rm{exp}[{\tau_{\rm{IGM}}}]$ term. For simplicity, we adopt a transmission value of 0.63, which is the mean IGM transmission value in the F154W band for a sample of four objects at redshifts $\sim 1.2$ in \cite{Dhiwar24}. The above value is obtained after averaging over 10000 lines of sight generated for each object, using the column density distribution prescription of HI absorbers from \cite{Inoue14}. Further details of the IGM models used can be found in \cite{Dhiwar24}. Thus, in the first approach, we proceed with estimating the escape fraction as, 

\begin{equation} 
  f^{\text{H}\alpha, \text{case B}}_{\rm{esc,LyC}}=\frac{\dot{N}^{esc}_{\rm{LyC}}}{\dot{N}^{esc}_{\rm{LyC}}+\dot{N}^{non-esc}_{\rm{LyC}, \text{case B}}}.
  \label{eq:fesc}
\end{equation}

The above definition reflects the standard definition of the absolute escape fraction with the caveat that it considers only those LyC photons that fall within the F154W band and is thus a lower limit to the absolute escape fraction. 
In addition to calculating the $f^{\text{H}\alpha, \text{case B}}_{\rm{esc,LyC}}$ above, that incorporates the IGM correction factor, we also estimate the escape fraction for the case of a fully transparent IGM line of sight. We denote this by $\tilde{f}^{\text{H}\alpha, \text{case B}}_{\rm{esc,LyC}}$. We provide our estimates in Table \ref{tab:galaxy10_prop}. Like for $\xi_{\rm{ion}},$ we also provide estimates for $f^{\text{H}\alpha, \text{case A}}_{\rm{esc,LyC}}$.
\par The second approach we adopt is to estimate $f_{\rm{esc}}$ from the SED fitting, which we denote as $f_{\rm{esc},\text{SED}}$. As mentioned in Section \ref{sec:sed} the escape of LyC photons is attributed only to G1. We note that the treatment of the IGM in CIGALE is adopted from the prescriptions of \citet{Meiksin06}. It is also worth highlighting that at wavelengths below the Lyman limit ($\lambda_{\rm{rest}}<912 \text{\AA}$), CIGALE does not implement the dust attenuation of the intrinsic spectra (see \cite{Kerutt23}).
Our $f_{\rm{esc},\text{SED}}$ estimates are provided in Table \ref{tab:galaxy10_prop}. We discuss our estimated escape fraction values in Section \ref{sec:discussion}.

\subsection{{Geometry of LyC escape }} \label{sec:geometry}
{Based on our estimated quantities and observables, we now attempt to predict the mechanism of LyC escape from the system. In a density bounded escape scenario, the partially ionized/neutral HI shell surrounding the HII regions is weak or perhaps absent leading to a low optical depth that facilitates LyC escape \citep{Zackrisson13,Paswanetal2022}. \cite{Wang19} propose that density bounded LyC escape can be traced by quantifying the relative weakness of  the [S II] 6717, 6731 emission lines, since the [S II] emission with a relatively low ionization potential of 10.4 eV arises from the outer edges of the Str{\"o}mgren sphere. They quantify the [S II] deficiency based on a sample of star-forming galaxies in the plane of $\rm{[S\: II]\lambda \lambda 6717,6731}/\rm{H}\alpha$ versus $\rm{[O\: III]}\lambda 5007/\rm{H}\beta$. We highlight this curve in green, in the right panel of Figure \ref{fig:galaxy10_grism1d}. The deficiency is defined as the horizontal displacement from this curve. Based on the emission line fluxes derived from the grism spectra, we cannot argue in favour of a deficiency in [S II].}
\par {We now examine the picket-fence model as a second possible geometry that leads to LyC escape from the system. In this model the LyC photons escape through low-optical depth HI channels in an otherwise neutral HI medium. This model is parameterized by the covering fraction of neutral hydrogen, $f_{\rm{cov}}(\text{HI})$. The dust content in the system has implications for the leakage of the LyC photons in this model, given that the HI and dust coexist in the star-forming regions of galaxies. It has however been posited with the support of simulations that LyC photons that escape a galaxy do so through channels that are also free of dust (see \cite{Reddy16,Kerutt23}). In light of this, in the following analysis, we assume that the channels through which the LyC photons escape are  dust-free. A model with this geometry has been previously adopted in the literature (e.g., \cite{Reddy16, Gazagnes18, Steidel18}). Following these works, we can write the escape fraction as,} 
\begin{equation} \label{eq:fesc_geometry}
  \begin{aligned}
    f_{\rm{esc},\text{LyC}} & = (1-f_{\rm{cov}}(\text{HI})) \: + \\
      & f_{\rm{cov}}(\text{HI}) \times \rm{exp}[-\tau^{\rm{cov}}_{\rm{ISM}}  (\rm{LyC})]\times 10^{-0.4\rm{k}(\rm{LyC})\rm{E(B-V)}_{*}},
      \end{aligned}
\end{equation}
{where $\tau^{\rm{cov}}_{\rm{ISM}}(\rm{LyC})$ denotes the optical depth due to photoelectric absorption, $\rm{k}(\rm{LyC})$ denotes the extinction factor for the dust attenuation curve at LyC wavelengths and $\rm{E(B-V)}_{*}$ is the continuum stellar reddening. We now approximate $f_{\rm{cov}}(\text{HI})$ by adopting the proposed empirical relation between $f_{\rm{cov}}(\text{HI})$ and $\rm{E(B-V)}_{*}$ by \cite{Reddy16}. The functional form of this relation is given by,
\begin{equation} \label{eq:cov_frac_EBV}
    f_{\rm{cov}}(\text{HI})=1-\rm{exp}[a \times \rm{E(B-V)}_{*}^{b} ],
\end{equation}
where, a,b are the fitted parameters.
To this end, we first use our estimate of the nebular reddening $\rm{E(B-V)}_{\text{neb}}$ from Section \ref{sec:dust}, to estimate the continuum reddening $\rm{E(B-V)}_{*}$, using the conventional conversion of $\rm{E(B-V)}_{*}=0.44 \times \rm{E(B-V)}_{\text{neb}}$ \citep{Calzetti00}. We obtain $\rm{E(B-V)}_{*} \sim 0.24$. We note that in the literature (e.g., \cite{Bassett19}), the relation $\rm{E(B-V)}_{*}=\rm{E(B-V)}_{\text{neb}}$ is also sometimes used, especially in the cases of high redshift galaxies ($z \gtrsim 2$).  Adopting the fitted values of a,b along with the corresponding errors from \cite{Reddy16}, we subsequently obtain $f_{\rm{cov}}(\text{HI}) \sim 0.96^{+0.01}_{-0.02} $.  In the limit that the second term of Equation \ref{eq:fesc_geometry} goes to zero, a reasonable assumption for column densities of the order $\rm{log}[\rm{N}(HI)/\rm{cm}^{-2}] \gtrsim 20$, the LyC escape fraction can be simply approximated, $ f_{\rm{esc},\text{LyC}} \approx 1-f_{\rm{cov}}(\text{HI})$ \citep{Reddy16}. Under this approximation, we estimate $f_{\rm{esc},\text{LyC}}\sim 0.04^{+0.02}_{-0.01}$. Our case B $f^{\rm{H}\alpha}_{\rm{esc},\text{LyC}}$ values and $f_{\rm{esc},\text{SED}}$ are larger than this estimate.} 
\par {We caution however that the above methodology may be too simplistic. Equation \ref{eq:fesc_geometry} may not fully capture the complex geometry that leads to LyC escape. Alternative models of dust-gas composition, such as a uniform dust-screen, may be applicable (e.g., \cite{SaldanaLopez22}). Furthermore, Equation \ref{eq:cov_frac_EBV} should ideally be applied only for averaged quantities over an ensemble of galaxies (see \cite{Reddy16, Steidel18}). For instance, the $\rm{E(B-V)}_{*}$ ($\sim 0.13-0.36$) and $f_{\rm{esc,LyC}}$ ($\sim 0.01-0.13$) values for a sample of low redshift LyC leakers studied by \cite{Gazagnes18} are comparable to the values in this work. The covering fractions estimated by \cite{Gazagnes18} using the Lyman series absorption lines however indicate that the individual LyC leakers have lower $f_{\rm{cov}}(\text{HI})$ (0.57-0.83) than predicted using Equation \ref{eq:cov_frac_EBV}. The strongest LyC leaker in the sample of \cite{Bassett19} ($f_{\rm{esc},\text{LyC}} \sim 0.4$) also has a moderate $\rm{E(B-V)}_{*}\sim 0.12$. We return to this particular leaker in Section \ref{sec:discussion}. Simulations of LyC leaking galaxies have shown that for individual galaxies, the escape fraction can have a significant line-of-sight dependence. For individual galaxies in such simulations, the line-of-sight $ f_{\rm{esc},\text{LyC}}$ can be either an order of magnitude above or below the global $ f_{\rm{esc},\text{LyC}}$, with recent results indicating that the latter case is more probable \citep{Choustikov24}.}
\section{Discussion and conclusions} 
\label{sec:discussion}
Our escape fraction estimates (case B) from both the approaches we use in this work are consistent with the range of values of $7-20 \%$ required for the completion of reionization by $z\sim 6$ \citep{Robertson13,Rosdahl18}. The ionizing photon production efficiency of G1 lies within the canonical values, $\text{log}(\xi_{\rm{ion}}/\text{Hz\: erg}^{-1})\simeq 25.2-25.3$ used in reionization scenarios \cite{Robertson13} and lies above the mean predicted value of $\xi_{\rm{ion}}$ at $z\sim 1$ \citep{Matthee17}. 
\par The LyC detection hints towards the presence of low column-density neutral hydrogen channels within the system. In the case of Haro 11, 21cm HI imaging has provided support for such a scenario. \cite{LeReste23} report a bulk offset of neutral hydrogen in Haro 11, likely caused due to the merger-driven interaction. While the observational evidence from recent studies points towards the increasing contribution of major mergers to the mass assembly of galaxies \citep{Duncan19}, whether these play a key role in reionization requires a careful investigation. \cite{Duncan19} find that for the stellar-mass range $\text{log}_{10}(\text{M}_{*}/\text{M}_{\odot})>10.3$, the major merger pair fraction increases almost linearly $\propto(1+z)$ and that the major merger rate per galaxy increases from $ 0.07\pm 0.01 \: \text{Gyr}^{-1}$ at $z<1$ to $7.6 \pm 2.7\: \text{Gyr}^{-1}$ at $z=6$, a period which is relevant for the EOR. Hydrodynamic simulations of mergers highlight the role played by the mass ratio of mergers in inducing star-formation within the system, with major mergers known to trigger starbursts \citep{Barnes91,Teyssier10,Rodriguez19}.
In this context, investigating the frequency of major mergers in galaxies with stellar masses below AF13753, alongside their LyC leakage properties, is important since low-mass galaxies are believed to dominate the ionizing budget. For a sample of low-mass ($\text{log}_{10}(\text{M}_{*}/\text{M}_{\odot})\lesssim 9.5$) Ly$\alpha$ emitting galaxies located within the EOR, \cite{Witten24} with the support of hydrodynamical simulations find evidence of mergers facilitating the escape of Ly$\alpha$ photons from these galaxies. They report that this is partly because of the increased intrinsic ionizing photons being produced within these systems during episodes of starbursts. Since the escape fraction from a system is time-dependent, one might also like to correlate the LyC escape with the merger stage of systems to understand the dominant cause of LyC leakage from these systems. 
\par In light of the analysis in Section \ref{sec:geometry}, the absence of evidence of density bounded regions and the relatively large dust content of AF13753 points towards the LyC photons escaping along the line of sight through a narrow cone in the ISM. 
This scenario is akin to those described in \cite{Bassett19} and more recently in \cite{Gupta24}. The analysis of the possible role of merger driven interactions in shaping the geometry would be an interesting future exercise. 
\par A high value ($>3$) of the O32 ratio is thought to be a necessary condition for LyC escape \citep{Nakajima20,Flury22}. In the star-forming knots of Haro 11, the O32 value exceeds 9 \citep{Keenan17}. We compare the extinction-corrected O32 value of AF13753 with some of the other known LyC candidate leakers for which this value is available, namely from the Low-redshift Lyman Continuum Survey ($z\sim0.2-0.4$, \cite{Flury22a}) and the sample of \cite{Bassett19} ($z\sim3.1-3.2$). 
 
Considering the empirical relation between $f_{\rm{esc,LyC}}$ and the O32 proposed by \cite{Izotov18}, our extinction corrected O32 value of 0.54 implies an $f_{\rm{esc,LyC}}$ value that lies an order of magnitude below our $f_{\rm{esc,LyC}}$ estimates. 37 out of the 66 galaxies in the parent sample of \cite{Flury22a} have been selected on the basis of their high O32 ($>3$). While nearly half of the LyC leakers of their sample have $\text{O}32>10$, \cite{Flury22} do find LyC leakers with escape fractions that lie considerably above the predicted O32-$f_{\rm{esc,LyC}}$ relation. These galaxies are found to have larger stellar masses ($>10^{9} \text{M}_{\odot}$) than the average of the sample. Interestingly, in the sample of \cite{Bassett19}, the galaxy with the highest estimated escape fraction ($\sim 0.4$), is found to have the lowest O32 ratio ($\sim 0.2$) and is the most massive ($\sim10^{10} \text{M}_{\odot}$). This galaxy shows signatures of a merger. The reduced O32 ratio in AF13753 and the strongest LyC leaker of \cite{Bassett19} could be explained by shocks, that are known to be common in ongoing mergers and can enhance the [O II] emission relative to the [O III] \citep{Epinat18}. 
Extreme emission-line galaxies, thought to be analogs of LyC leakers in the EOR are characterized by large H$\beta$ and [O III] equivalent widths \citep{Endsley21}.
\cite{Flury22} find that the majority of strong LyC leakers (non-detections) in their sample have an H$\beta$ EW $>150 \text{\AA}$ ($<150 \text{\AA}$). Their sample does however contain a few strong LyC leakers with escape fractions ($\sim 0.1$) and H$\beta$ EW ($\lesssim 50 \text{\AA}$) comparable to AF13753. 
\par Systems like AF13753 may thus hint towards a sub-population of LyC leakers which do not fall under the conventional selection strategy adopted in such studies. 
However, we caution against an overinterpretation of such a scenario keeping in mind the low-number statistics. The resolution of UVIT allows us to investigate only a limited number of interacting candidates to test if they are LyC leakers. An investigation of merging candidates at different redshifts for which one can trace the LyC using higher-resolution imaging could be beneficial. We defer a more careful examination of such statistics for future work.  
\par Our main findings in this work are summarised below:
\begin{itemize}
    \item Combining high-resolution multiband HST and JWST images with UVIT far-ultraviolet imaging, we confirm the detection of LyC photons from a likely massive major merger interacting system, AF13753 at $z\sim 1.1$.
    \item We report an infrared luminous galaxy (G2) in the system that is faint in the rest-frame ultraviolet and optical. This galaxy lies roughly 5.5 kpc away from the dominant site of star-formation in the system. We suspect that the disturbed morphology and the complex ionized-gas kinematics of the system are shaped by the merger driven interactions. Our SED analysis of the two galaxies suggests that this is a massive, dust-rich system. The analysis favours G1 having a higher star-formation rate and lower metallicity than G2.
   
    \item We favour an interpretation of the LyC photons escaping G1, through a narrow low HI optical depth dust-free channel. The anisotropic escape may make the detection of such systems difficult. 
    \item We constrain the escape fraction of G1 to be greater than around eight percent and the ionizing photon production efficiency to be around 25.45, assuming the standard case B recombination.  Conventional LyC diagnostics do not favour AF13753 being a LyC leaker. Thus, the non-negligible $f_{\rm{esc,LyC}}$ and $\xi_{\rm{ion}}$ values of AF13753, provides motivation for a future investigation into the potential significance of merging systems to reionization.
\end{itemize}

Some of the data presented in this article were obtained from the Mikulski Archive for Space Telescopes (MAST) at the Space Telescope Science Institute. The specific observations analyzed can be accessed via the following \dataset[DOI]{https://doi.org/10.17909/ms43-d869}.
\smallskip \\
We thank the anonymous referee for their careful review and suggestions that have improved the quality of this manuscript.

\facilities{AstroSat (UVIT), HST (WFC3, ACS-WFC, WFC3-UVIS), JWST (NIRCam), VLT (MUSE)}


\software{astropy \citep{2013A&A...558A..33A,2018AJ....156..123A},  
         Source Extractor \citep{Bertin96}, \texttt{grizli} (\href{https://github.com/gbrammer/grizli/}{https://github.com/gbrammer/grizli/}), \textit{Photutils} \citep{Bradley23}, CIGALE \citep{Boquien19} 
          }


\appendix
\section{Examples of discarded candidates} \label{sec:eg_discard}
\begin{figure}[ht!]
\nolinenumbers
\centering
\vspace{0.6cm}
\includegraphics[width=0.8\columnwidth]{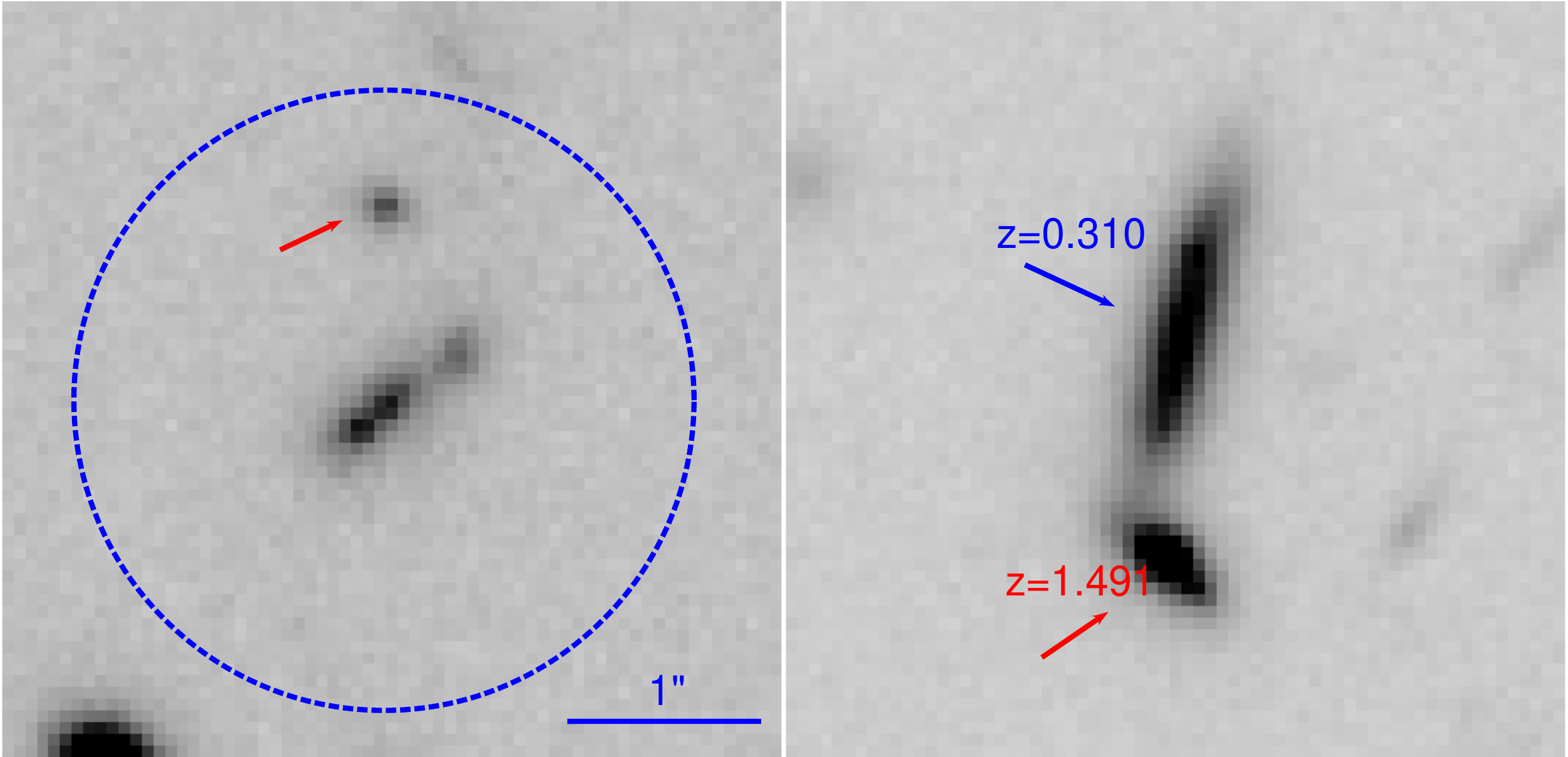}
\caption{Examples of discarded candidates based on our selection criteria. The band used for this is the HST F850LP. In the left panel the merging candidate is enclosed within the aperture of radius 1.6" which is the PSF FWHM of the UVIT F154W band. The object indicated by the red arrow is a possible contaminant. In the right panel we highlight an apparent interacting system that is actually comprised of two objects at different spectroscopic redshifts \citep{Bacon23}.} 
\label{fig:contameg}
\end{figure}

In this section we highlight two examples of candidates that do not meet our required selection criterion in Section \ref{sec:data_sample}. In Figure \ref{fig:contameg}, the left panel highlights a  merger candidate in the HST F850LP band, located at the centre of the aperture which is of radius 1.6". We indicate the possible contaminant in this panel by the red arrow. This object does not show signs of interaction with the candidate merger system and we do not have its spectroscopic redshift. Thus, we reject this candidate merger.  In the right panel, we highlight a case in which we leverage the available MUSE spectroscopic coverage in the field to reject an apparent merger system identified in the previous steps. The different redshifts of the apparent interacting components rule out this object as a merger. In our selection criterion at this step, seven out of the eleven discarded candidates fall under the case corresponding to the left panel and the rest fall under the case corresponding to the right panel. 

\section{MUSE 1D spectra} \label{sec:muse1d}
\begin{figure*}[ht!]
\nolinenumbers
\includegraphics[width=1\textwidth]
{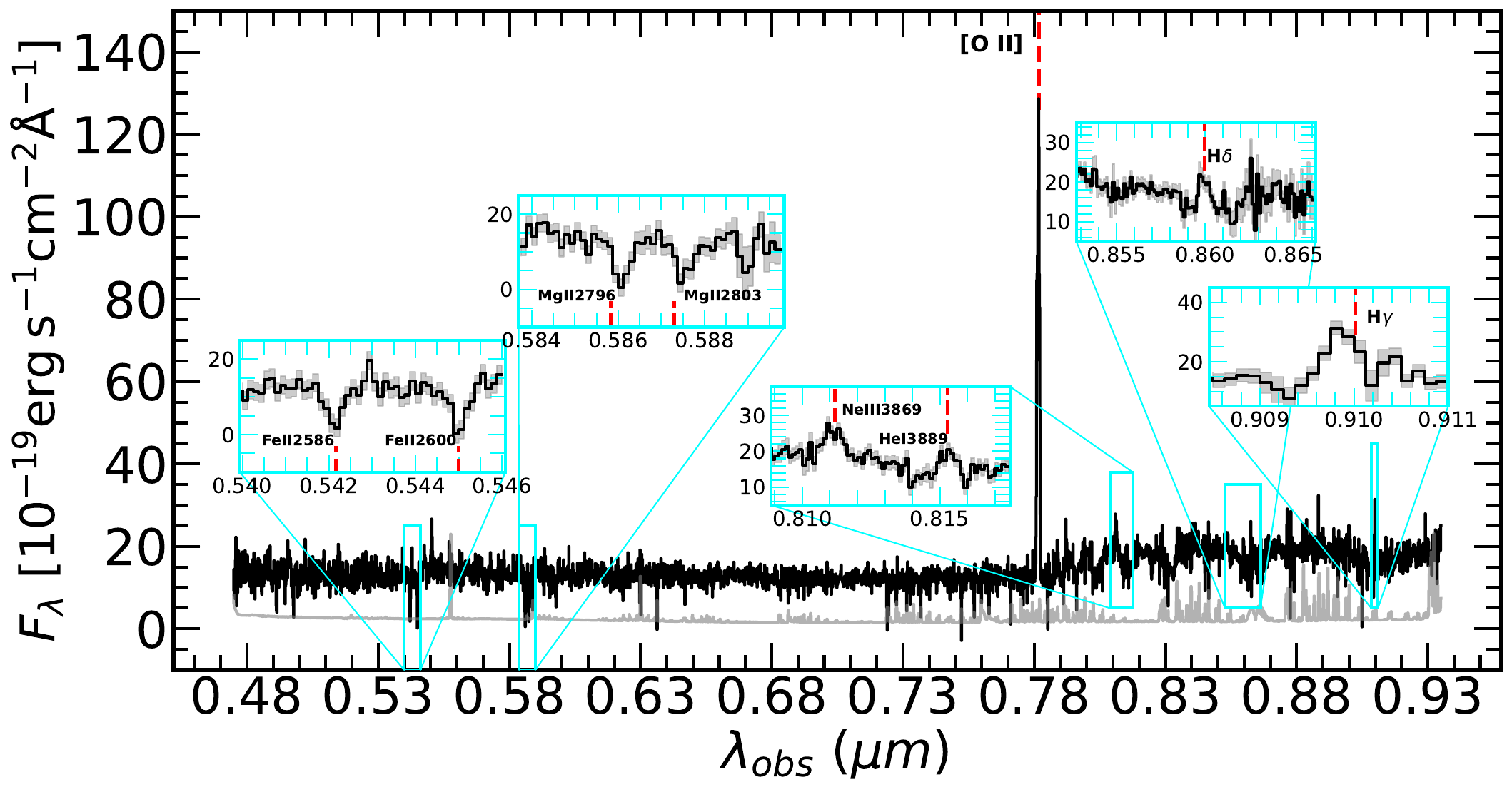} 
\caption{The black curve denotes the 1D Muse spectrum of AF13753. Other than the [O II] doublet (indicated by the red-dashed line), we highlight all the emission and absorption lines we identify in the inset plots. The grey spectrum denotes the error spectrum. For the inset plots, we display the error spectrum as an envelope around the data.}

\label{fig:galaxy10_MUSE_1D}
\end{figure*}

The integrated MUSE 1D spectra of AF13753 (MUSE ID: 899) is extracted from the MUSE data cube of the HUDF using an aperture of radius 1.6" paced at the coordinates mentioned in Table \ref{tab:galaxy10_prop}. All of the identified lines for this object in the plot are present in the catalog of \cite{Bacon23}. We do not report any unidentified emission lines in the spectra (Figure \ref{fig:galaxy10_MUSE_1D}) which could indicate a redshift solution other than 1.097.
\par
\begin{figure*}[ht!]
\nolinenumbers
\includegraphics[width=0.75\textwidth]
{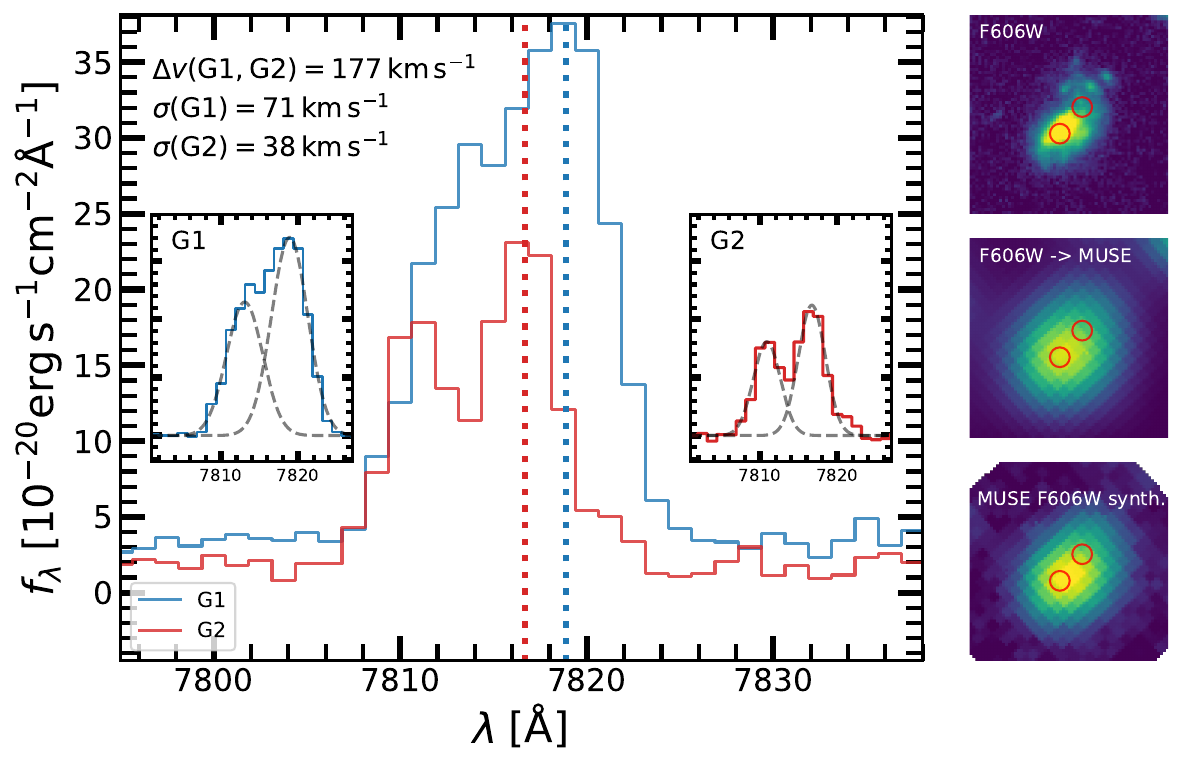}
\centering
\caption{The 1D [O II] $\lambda \lambda 3727, 3729$ doublet emission lines extracted from the MUSE data cube at the location of G1 and G2. The blue spectrum and red spectrum is extracted from the location of G1 and G2 respectively. The vertical dotted lines indicate the peak of the fitted [O II]$\lambda \lambda$ 3729 line. The inset plots display the fitted models to the [O II] doublet indicated by the grey dashed curves. The velocity difference between the two spectra and their individual dispersions are indicated in the top left of the plot. On the right: F606W image (top), HST F606W image convolved with the MUSE PSF and resampled to MUSE pixel scale (middle), and the MUSE synthesized F606W image (bottom). The red circles denote apertures of radius 0.2" (single pixel) used to extract the spectra plotted in the left. North is up and East is to the left in all the panels as in Figure \ref{fig:galaxy10_images}.}
\label{fig:galaxy10_OII_v2}
\end{figure*}

 The 1D spectra highlighting just the [O II] emission in Figure \ref{fig:galaxy10_OII_v2} are extracted using apertures of radius single pixel ($0.2"$) in order to minimize the mutual flux contamination. 
The MUSE synthesised image is constructed by convolving the image taken by the MUSE instrument with the HST F606W filter transmission curve. For comparison, we directly convolve the HST F606W image with the MUSE PSF and resample the image to the MUSE pixel scale. We do not find significant differences between the two images, indicating that the MUSE data is sensitive to the combined light of G1 and G2.

\bibliography{lyc_interacting_ApJ_arxiv}{}
\bibliographystyle{aasjournal}



\end{document}